\documentclass[pra,showpacs,superscriptaddress,twocolumn]{revtex4-2}

\usepackage[dvipdfmx]{graphicx}
\usepackage{amsmath}
 
\graphicspath{{graphics/}}  
\usepackage{color}
\usepackage{cancel}
\usepackage{ulem}
\usepackage{comment} 
\usepackage{bm} 
\usepackage{braket}
\usepackage{amsmath}
\usepackage{amssymb}

\usepackage{float}

\begin{document}
\title{Optical vortex generation by magnons with spin-orbit-coupled light}

\author{Ryusuke~Hisatomi}
\email{hisatomi.ryusuke.2a@kyoto-u.ac.jp}
\affiliation{Institute for Chemical Research (ICR), Kyoto University, Gokasho, Uji, Kyoto 611-0011, Japan}
\affiliation{Center for Spintronics Research Network (CSRN), Kyoto University, Gokasho, Uji, Kyoto 611-0011, Japan}
\author{Alto~Osada}
\affiliation{Center for Quantum Information and Quantum Biology (QIQB), Osaka University, Toyonaka, Osaka 560-0043, Japan}
\author{Kotaro~Taga}
\affiliation{Institute for Chemical Research (ICR), Kyoto University, Gokasho, Uji, Kyoto 611-0011, Japan}
\author{Haruka~Komiyama}
\affiliation{Institute for Chemical Research (ICR), Kyoto University, Gokasho, Uji, Kyoto 611-0011, Japan}
\author{Takuya~Takahashi}
\affiliation{Institute for Chemical Research (ICR), Kyoto University, Gokasho, Uji, Kyoto 611-0011, Japan}
\author{Shutaro~Karube}
\affiliation{Institute for Chemical Research (ICR), Kyoto University, Gokasho, Uji, Kyoto 611-0011, Japan}
\affiliation{Center for Spintronics Research Network (CSRN), Kyoto University, Gokasho, Uji, Kyoto 611-0011, Japan}
\author{Yoichi~Shiota}
\affiliation{Institute for Chemical Research (ICR), Kyoto University, Gokasho, Uji, Kyoto 611-0011, Japan}
\affiliation{Center for Spintronics Research Network (CSRN), Kyoto University, Gokasho, Uji, Kyoto 611-0011, Japan}
\author{Teruo~Ono}
\affiliation{Institute for Chemical Research (ICR), Kyoto University, Gokasho, Uji, Kyoto 611-0011, Japan}
\affiliation{Center for Spintronics Research Network (CSRN), Kyoto University, Gokasho, Uji, Kyoto 611-0011, Japan}
\affiliation{International Center for Synchrotron Radiation Innovation Smart, Tohoku University, Sendai, Miyagi 980-8577, Japan}

\date{\today}

\begin{abstract}
Light possesses both spin and orbital angular momentum. In spatially inhomogeneous optical fields, which are characterized by a finite distribution of wave vectors in $k$ space, these properties undergo spontaneous coupling, referred to as optical spin-orbit coupling. The study of the coupling has recently become central in modern optics due to its substantial applications in communications, sensing, and quantum control. A key challenge is to clarify the relationship between the origins of spatially inhomogeneous optical fields and the resulting spin-orbit coupling. Current research focuses on materials and configurations exhibiting spatial inhomogeneity, such as focusing lenses, interfaces, inhomogeneous media, and metasurfaces. However, Maxwell's equations indicate that matter can introduce both spatial inhomogeneity and temporal asymmetry into optical fields. For instance, magnetic ordering breaks the time-reversal symmetry of interacting optical fields via the magneto-optic effect, introducing nonreciprocity in the resulting optical phenomena. Despite the importance, optical phenomena involving both spatially inhomogeneous and temporally asymmetric optical fields remain unexplored. Here, we demonstrate that combining temporal asymmetry and spatial inhomogeneity through magnons and light focusing, respectively, transforms an input Gaussian beam into a specific optical vortex beam in a nonreciprocal manner. This phenomenon is quantitatively explained by integrating the physics of magnon-induced Brillouin light scattering with optical spin-orbit coupling. The observed conservation of total angular momentum, encompassing both magnons and photons, further indicates that magnons can control both spin and orbital angular momentum of light. Finally, we outline future research directions enabled by optical fields exhibiting both spatial inhomogeneity and temporal asymmetry.
\end{abstract}


\maketitle
\section{Introduction}
Light carries both spin and orbital angular momentum (OAM)~\cite{LM1992}, which manifest as polarization and optical vortices, respectively. Traditional geometrical optics treats polarization and spatial characteristics as independent, neglecting their intrinsic coupling~\cite{YY1990}. However, Maxwell's equations inherently link optical fields to spatial variations---or equivalently, a finite distribution of wave vectors---making optical spin-orbit coupling~\cite{K2009,KE2011,KF2015} a fundamental property of all optical processes. Over the past quarter century, the exploration and classification of spin-orbit coupling have become central challenges in modern optics research, driven by significant applications in communications~\cite{LE2011,JJ2014}, sensing~\cite{OD2010,AA1997}, and quantum control~\cite{AA2024,CD2013}. A primary research objective is to elucidate the relationship between the origins of spatially inhomogeneous optical fields and the resulting spin-orbit coupling. The relationship between materials and structures exhibiting spatial inhomogeneity, such as focusing lenses~\cite{KE2011,C2009,PP2009,KM2010}, interfaces~\cite{KY2004,MS2004}, inhomogeneous media~\cite{K2009,VB1992,AG2003}, and metasurfaces~\cite{ZG2002,LX2013,NF2014}, and the spin-orbit-coupled light they generate has already been established.

\begin{figure*}[t]
\begin{center}
\includegraphics[width=18.0cm,angle=0]{Concept_240610_forPaper_W1800_v22_PRstyle.pdf}
\caption{
Optical vortex generation via magnon-induced Brillouin light scattering (BLS). (a) Schematics of optical vortex generation manifested when the direction of an external magnetic field applied to a ferromagnetic sphere and the direction of input and scattered light coincide. Inset shows the orbital angular momentum of Stokes and anti-Stokes scattered light. Note that this only shows the situation where the helicity of input and scattered light is the same (left circular), i.e., helicity-conserving BLS. These scatterings satisfy the conservation of total angular momentum. (b) Schematics of light propagation and the mechanism of optical vortex generation. Spin-orbit-coupled light within the sphere possesses an electric field in the $z$ direction (green arrows). Note that this illustrates the situation with the leftmost scattering in the inset of Fig.~\ref{fig:concept}(a), and the significant Stokes sideband with $\Delta l_p=+1$ in Fig.~\ref{fig:eff_H}(a). Transitions between spin-orbit-coupled-light modes within the sphere, which occur only when OAM is identical and magnons can mediate spin angular momentum (SAM) changes, generate optical vortices.}
\label{fig:concept}
\end{center}
\end{figure*}

\vspace{5mm}

Investigating the optical spin-orbit coupling in the presence of magnetic ordering offers interesting prospects, as the time-reversal symmetry breaking is intertwined with spatially inhomogeneous optical fields. For instance, the time-reversal symmetry breaking induced by magnetic ordering on interacting light could introduce nonreciprocity~\cite{L1931_1,L1931_2,CA2018} into spin-orbit coupling phenomena. Conversely, the optical spin-orbit coupling could offer new perspectives on magneto-optics, which has traditionally focused only on the spin angular momentum or polarization of light~\cite{P1963,P1967}. Furthermore, in ferromagnetic materials under a designed external magnetic field configuration, magnons with gigahertz-order resonance frequencies emerge~\cite{TH1940}, potentially enabling high-speed control of optical angular momentum for communication applications~\cite{AH2015,J2016,AK2021,JJ2022,CA2023}. Examining the interaction between magnons and spin-orbit-coupled light could provide new insights for topological photonics~\cite{MJ2014,LJ2014}, optomagnonics~\cite{BS2022}, and chiral quantum optics~\cite{PS2017}.

Driven by these prospects, we have constructed a system in which an external magnetic field is applied to a 0.5-mm-diameter ferromagnetic sphere, and light is transmitted parallel to the applied magnetic field. Here, we report the first experimental realization of the nonreciprocal transformation of an input Gaussian beam into a specific optical vortex beam, achieved through the combined temporal asymmetry and spatial inhomogeneity induced by magnons and light focusing, respectively. The geometric arrangement of the light and external magnetic field, as well as a representative example of optical vortex generation, is shown in Fig.~\ref{fig:concept}(a). The results confirm that the total angular momentum between photons and magnons is conserved in all observed phenomena.

\begin{figure*}[t]
\begin{center}
\includegraphics[width=18.0cm,angle=0]{Optical_Setup_240405_forPaper_W180_v15_W1800_PRstyle.pdf}
\caption{
Experimental setup and characterization of uniform magnon mode. (a) Experimental setup for free-space propagating part of light. A spherical monocrystal ($0.5~\rm{mm}$ diameter) of YIG is placed in the gap of a magnetic circuit consisting of a pair of cylindrical permanent magnets, a coil, and a yoke. The external magnetic field is parallel to the crystal axis $\langle100\rangle$ and $z$ axis as well. A coupling loop coil above the YIG sphere is used to excite magnons in uniform mode. Two sets of quarter wave plate (QWP) and polarizing beam splitter (PBS) are used to discriminate the helicity of the input and scattered light, respectively. Collimated-Gaussian-mode laser beam emitted from a single-mode (SM) fiber through a collimating lens (not shown) is focused to about $100\,\mathrm{\mu m}$ by a convex lens and injected into the YIG. The scattered light that comes out coaxially with the transmitted light is reflected by a spatial light modulator (SLM) and coupled to another SM fiber to identify the optical OAM. (b) Heterodyne measurement system. Light from a continuous-wave (CW) laser is separated into two paths by a fiber splitter. An electro-optic modulator (EOM) in the upper path is used to calibrate the signal, and an acousto-optic modulator (AOM) in the lower path is used to generate a local oscillator (LO). The signal and the LO are combined, and the resulting signal is sent to a high-speed photodetector (HPD), followed by a microwave amplifier, and then to a spectrum analyzer. (c) Schematics of the relevant frequencies. The carrier light at $\Omega_C$ is scattered into the sidebands at $\Omega_R$ and $\Omega_B$. The beat signals appear at $\omega_R$ and $\omega_B$. (d) The relationship between a liquid crystal pattern of the SLM set and a scattered optical vortex, which is to be converted to the fundamental Gaussian mode by the SLM. (e) Microwave reflection spectra $|\rm{S}_{11}|$ for the uniform magnon mode. The blue lines show the measured reflection amplitude and phase, whereas the red curves show the fitting results using an appropriate function~\cite{RA2016}.}
\label{fig:setup}
\end{center}
\end{figure*}

\section{Concept of optical vortex generation and detection }
We proceed with the experiment in the following procedure. First, a uniform magnetostatic mode, known as the Kittel mode~\cite{L1958,PI1959}, in a ferromagnetic spherical monocrystal (0.5 mm in diameter) of yttrium iron garnet (YIG) is coherently excited. A circularly polarized Gaussian beam with a diameter of 50 $\rm{\mu m}$ and a wavelength of 1.5 $\mathrm{\mu m}$ is incident onto the ferromagnetic sphere. This beam is gently focused, i.e., pseudo-paraxial light~\cite{ME2005}, using a convex lens with a focal length of 100 mm positioned in front of the sphere. Since both the incident and exit spherical surfaces behave as a convex lens with a focal length of about 230 $\mathrm{\mu m}$, under proper alignment such that the beam waist roughly coincides with the center of the sphere, the modeling shown in Fig.~\ref{fig:concept}(b) is approximately justified. Within the sphere, focusing and divergence occur, and the propagating light behaves as nonparaxial light. Subsequently, light scattered by uniform-mode magnons is reconverted to collimated light by multiple lenses. We then identify optical spin angular momentum (SAM, i.e., helicity) and OAM~\cite{LM1992} in the far-field region, determine whether the scattering is Stokes or anti-Stokes, and quantitatively evaluate the scattering efficiency for each scattering process. For details on these series of techniques, refer to Appendix~\ref{sec:methods_opto}.

The whole experimental setup is schematically shown in Figs.~\ref{fig:setup}(a) and \ref{fig:setup}(b). The YIG sphere is placed at the center of the gap of a magnetic circuit and saturated by applying a magnetic field of around $150\,\rm{kA/m}$ along the crystal axis $\langle100\rangle$ and $z$ axis. This magnetic ordering, determined by the external magnetic field, breaks the time-reversal symmetry of the whole system. A coupling loop coil above the YIG sphere generates an oscillating magnetic field perpendicular to the saturation magnetization to excite magnons in the uniform mode. Figure~\ref{fig:setup}(e) shows the microwave reflection spectra $|\rm{S}_{11}|$ acquired by a vector network analyzer, indicating the ferromagnetic resonance for the uniform mode. The resonance frequency of the mode, $\omega_m/2\pi=3.730\,\rm{GHz}$, and the number of excited magnons are estimated from the Lorentzian fitting~\cite{RA2016}.

For paraxial light in the far-field region, Laguerre-Gaussian $\left({\rm{LG}}_{p}^{l_p}\right)$ modes are a possible set of basis vectors~\cite{LM1992}. The index $l_p$ is a winding number, and $(p+1)$ is the number of radial nodes. In this paper, we only consider cases of $p=0$. 
This simplification is experimentally justified because the combination of an SLM and an SM fiber acts as a strict spatial filter that selectively couples only the $p=0$ mode, effectively suppressing any potential contamination from higher-order radial modes with $p \neq 0$ [see Fig.~\ref{fig:setup}(a) and Appendix~\ref{sec:methods_opto} for details].
The azimuthal phase term $\left(e^{il_p\phi}\right)$ of the LG modes results in a helical wave front, and therefore the light is called an optical vortex. The handedness of the helical wave fronts of the LG modes is linked to the sign of the index $l_p$. The OAM per photon of LG modes with respect to the propagation direction is $l_p \hbar$. On the other hand, the SAM per photon is $s_p \hbar$, where $s_p=\pm{1}$ for the left or right circularly polarized light. LG modes can simultaneously carry two angular momenta, SAM and OAM. In this paper, we denote optical modes as ${\rm{LG}}_{l_p,s_p}$ with OAM in the first subscript and SAM in the second subscript.

\begin{figure*}[t]
\begin{center}
\includegraphics[width=18.0cm,angle=0]{BarPlots_H_Kittel_100axis_W180_rev11_W1800_PRstyle.pdf}
\caption{
Scattering efficiencies. (a)-(d) Scattering efficiencies of the Stokes sideband (red bars) and the anti-Stokes sideband (blue bars) for four distinct optical polarization sets under the external magnetic field $\bm{H}\parallel\langle100\rangle$. The height of the color bar shows the mean scattering efficiency, and the difference between the top of the black wireframe and the bar represents a standard deviation estimated from measurements repeated six times. The gray arrows indicate scattering that satisfies the conservation of total angular momentum. Note that the sign of each angular momentum is defined by a quantization axis oriented in the positive direction of the $z$ axis.
}
\label{fig:eff_H}
\end{center}
\end{figure*}

\section{Experiments}
\subsection{Optical vortex generation by magnons}
Figures~\ref{fig:eff_H}(a)-\ref{fig:eff_H}(d) show the observed magnon-induced BLS efficiencies. The scattering efficiencies, which are probabilities that one magnon scatters incident one photon, are deduced from the signal at the angular frequency of $\omega_R=\omega_m+\omega_A$ for the Stokes scattering and that at $\omega_B=\omega_m-\omega_A$ for the anti-Stokes scattering. The calibration scheme is provided in the supplemental information in Ref.~\cite{RA2019}. Note that no signal is produced when the scattered light is intercepted before the single-mode fiber on the output in Fig.~\ref{fig:setup}(a), confirming that no stray signal is directly coupled to the high-speed photodetector. The sign and magnitude of the SAM $s_p$ per input or output photon~\cite{LM1992} and SAM $s_m$ per magnon~\cite{JJ1958} are determined from the definitions and experimental configuration. The OAM of scattered photon, $l_s$, is determined using a method (see Appendix~\ref{sec:methods_opto}). Let $\Delta s_p$, $\Delta s_m$, and $\Delta l_p$ denote the change in each angular momentum in each scattering. Since the input Gaussian beam has no OAM, $\Delta l_p$ and $l_s$ are identical, and the fact that $\Delta l_p$ is nonzero means that the scattered photon is an optical vortex.

In the case where the input and output polarization are both left circular ($L_i\rightarrow L_o$ configuration) shown in Fig.~\ref{fig:eff_H}(a), the significant Stokes sideband with $\Delta l_p=+1$ (red bar indicated by gray arrow) appears with a scattering efficiency of $0.85\times10^{-22}$. On the other hand, the significant anti-Stokes sideband with $\Delta l_p=-1$ (blue bar indicated by gray arrow) also appears with a scattering efficiency of $0.89\times10^{-22}$. Note that, from here on, we discuss the OAM as an integer value in units of $\hbar$. The results show that a vortex-free Gaussian beam ($l_p=0$) becomes a superposition of two optical vortex beams with different frequencies and OAM via the BLS induced by vortex-free magnon, as shown in Fig.~\ref{fig:concept}(a), which has long been unnoticed. To understand the rule that governs the observed scattering, we now focus on the elementary processes of magnon-induced BLS~\cite{YN1965}. Since the process can be viewed as a three-wave mixing process involving one magnon and two photons, the change in total angular momentum in the BLS can be written as $\Delta s_m+\Delta s_p +\Delta l_p$. We can verify that the observed helicity-conserving scattering processes in Fig.~\ref{fig:eff_H}(a) satisfy total angular momentum conservation, 
\begin{equation}
    \Delta s_m + \Delta s_p + \Delta l_p =0. \label{eq:conservation}
 \end{equation}
In this helicity-conserving scattering, angular momenta exchange only between magnon SAM and photon OAM.

We now focus on the helicity-changing scattering. For the case where the input (output) polarization is left (right) circular ($L_i \rightarrow R_o$ configuration) in Fig.~\ref{fig:eff_H}(b), the only significant anti-Stokes sideband with $\Delta l_p=+1$ (blue bar indicated by gray arrow) appears with a scattering efficiency of $0.53\times10^{-22}$. For the case where the input (output) polarization is right (left) circular ($R_i \rightarrow L_o$ configuration) in Fig.~\ref{fig:eff_H}(c), the only significant Stokes sideband with $\Delta l_p =-1$ (red bar indicated by gray arrow) appears with a scattering efficiency of $0.57\times10^{-22}$. These results also show that scattering processes that conserve total angular momentum in Eq. (1) are allowed. It should be noted that the scattering efficiencies of allowed processes in Figs.~\ref{fig:eff_H}(a)-\ref{fig:eff_H}(c) are different. The scattering efficiencies of the helicity-conserving scattering in Fig.~\ref{fig:eff_H}(a) are approximately $2\,\rm{dB}$ higher than those of the helicity-changing scattering in Figs.~\ref{fig:eff_H}(b) and \ref{fig:eff_H}(c).

The lowest scattering efficiency is observed when the incident and output polarization are right-handed circular ($R_i\rightarrow R_o$ configuration), as shown in Fig.~\ref{fig:eff_H}(d). The slight Stokes sideband with $\Delta l_p=+1$ (red bar indicated by gray arrow) appears with a scattering efficiency of $0.07\times 10^{-22}$. On the other hand, the slight anti-Stokes sideband with $\Delta l_p=-1$ (blue bar indicated by gray arrow) appears with a scattering efficiency of $0.07\times10^{-22}$. The two scattering processes indicated by the gray arrows in Fig.~\ref{fig:eff_H}(d) satisfy total angular momentum conservation, although they are approximately $11\,\rm{dB}$ smaller than the results in Fig.~\ref{fig:eff_H}(a).

\subsection{Nonreciprocal optical vortex generation by magnons}
To get further insight, we repeat the same experiment with only the external magnetic field direction reversed. The results, shown in Fig.~\ref{fig:eff_invH}, demonstrate that scattering processes that satisfy the angular momentum conservation in Eq.~(\ref{eq:conservation}) occur and that the magnitude of the scattering efficiency depends on the direction of the magnetic field. Specifically, the order of the magnitude of the efficiencies of allowed scattering processes in Figs.~\ref{fig:eff_invH}(a)-\ref{fig:eff_invH}(d) is the reverse of the result in Figs.~\ref{fig:eff_H}(a)-\ref{fig:eff_H}(d). Additionally, comparison of Figs.~\ref{fig:eff_H} and \ref{fig:eff_invH} confirms the nonreciprocal nature of this phenomenon.

\section{Theory}
The key to quantitatively understanding the observed scattering lies in the optical spin-orbit interaction acting concurrently with the magneto-optic effect. The input-paraxial-light mode from outside the YIG sphere is transformed into superposition of spin-orbit-coupled-light (non-paraxial-light) modes through refraction at the spherical surface, as~\cite{KE2011}
\begin{eqnarray}
{\rm{LG}}_{l_p,s_p}=&&\sqrt{\cos \theta}[a\times{\rm{LG}}_{l_p,s_p}^{\rm{SO}} \notag\\
&&-b\times{\rm{LG}}_{l_p+2s_p,-s_p}^{\rm{SO}}-\sqrt{2ab}\times{\rm{LG}}_{l_p+s_p,0}^{\rm{SO}}], \label{eq:OSOI_main}
\end{eqnarray}
where $a=\cos^2(\theta/2)$, $b=\sin^2(\theta/2)$, and $\theta$ is an aperture angle defined in Fig.~\ref{fig:concept}(b), which in our system is evaluated as $\theta = \tan^{-1}(25\ \mathrm{\mu m} / 230\ \mathrm{\mu m}) \approx 0.11\ \mathrm{rad}$ from the incident Gaussian beam radius ($25\ \mathrm{\mu m}$) at the sphere surface and the focal length ($230\ \mathrm{\mu m}$). 
Physically, this transformation into spin-orbit-coupled modes originates from the spatial inhomogeneity of the light field, characterized by a finite distribution of wave vectors in $k$ space that deviates from the paraxial limit. While the incident Gaussian beam is quasi paraxial and exhibits negligible spin-orbit interaction, the lensing effect of the YIG sphere broadens the wave vector distribution, making these spin-orbit-coupled components—represented by the second and third terms in Eq.~(\ref{eq:OSOI_main})—essential for describing the internal field.
In each mode on the right-hand side of Eq.~(\ref{eq:OSOI_main}), the original total angular momentum $l_p+s_p$ is conserved. The third term represents an optical vortex state with a longitudinal electric field. As $\theta$ increases, the third term becomes significant, though it is one order of magnitude smaller than the first term. The second term is one order of magnitude smaller than the third term and is typically negligible. When the spin-orbit-coupled light refracts out of the sphere again without interacting with the magnons, it returns to the original-paraxial-light mode ${\rm{LG}}_{l_p,s_p}$.

To understand the observed optical vortex generation, it is necessary to consider the magneto-optic effect among spin-orbit-coupled modes $\rm{LG}^{SO}$ within the sphere, rather than the magneto-optic effect between far fields described by existing theories. Namely, the magneto-optic effect by uniform-mode magnons possessing only SAM $|s_m|=1$ allows transitions between the spin-orbit-coupled modes sharing the same OAM but differing by SAM of $1$. As described in Fig.~\ref{fig:concept}(b), taking the Stokes sideband at $\omega_R$ shown in Fig.~\ref{fig:eff_H}(a) as an example, scattering from $\rm{LG}_{0,+1}$ to ${\rm{LG}'}_{+1,+1}$ in the far field occurs due to a transition from ${\rm{LG}}_{+1,0}^{\rm{SO}}$ to ${\rm{LG}'}_{+1,+1}^{\rm{SO}}$ within the YIG sphere. Its efficiency is proportional to $\left[a\sqrt{2ab}\times\left(-f+2G_{44}\left(\frac{\mu_B n}{2}\right)\right)\right]^2$, where $f$ and $G_{44}$ are the Faraday coefficient and Cotton-Mouton coefficient of YIG, $\mu_B$ is Bohr magneton, $n$ is spin density of YIG, and the prime symbol indicates a scattered-light mode. The efficiency is indeed described by a combination of quantities related to the optical spin-orbit coupling and the magneto-optic effect. Note that the transition from ${\rm{LG}}_{+2,-1}^{\rm{SO}}$ to ${\rm{LG}'}_{+2,0}^{\rm{SO}}$ within the sphere also contributes, but it is negligible as it is 5 orders of magnitude less efficient than the preceding process (see Appendix~\ref{sec:details_theory}).

\section{Discussion and outlook}
The observed scatterings in Figs.~\ref{fig:eff_H}(a)-\ref{fig:eff_H}(d) and \ref{fig:eff_invH}(a)-\ref{fig:eff_invH}(d) are reproduced entirely by our theoretical framework outlined in the above section. First, the polarization-dependent efficiency differences observed experimentally—$2.3\,\rm{dB}$ [between anti-Stokes scatterings in Figs.~\ref{fig:eff_H}(a) and \ref{fig:eff_H}(b)] and $11\,\rm{dB}$ [between anti-Stokes scatterings in Figs.~\ref{fig:eff_H}(a) and~\ref{fig:eff_H}(d)]—are found to agree well with theoretical values derived from literature and known values, which are $2.7\,\rm{dB}$ and $11\,\rm{dB}$, respectively (see Table~\ref{table:comparison_exp_theo}). These differences are attributed to the relative sign difference between the Faraday and the Cotton-Mouton coefficients, as well as the superposition of multiple scattering processes. Furthermore, the theory reproduces the absolute values of efficiencies. Taking the anti-Stokes sideband with $\Delta l_p=-1$ [blue bar indicated by gray arrow in Fig.~\ref{fig:eff_H}(a)] as an example, we can confirm the orders of magnitude of experimental and theoretical values: The experimental value is $0.89\times10^{-22}$ and the theoretical value is $1.4\times 10^{-22}$ (see Appendix~\ref{sec:details_theory} and Table~\ref{table:comparison_exp_theo}). For all other observed scattering efficiencies, the theoretical and experimental values also agree well (see Table~\ref{table:comparison_exp_theo}).

Our theory also provides a reason for another crucial result. Table~\ref{table:AM} shows the possible angular momentum changes in this paper. From this, for instance, we can expect that the combination $(\Delta s_m,\Delta s_p, \Delta l_p)=(+1,+2,-3)$ satisfies the conservation of total angular momentum in Eq.~(\ref{eq:conservation}). However, as described in Fig.~\ref{fig:eff_H_LG03}, such scattering is not observed. Our theory supports the existence of this scattering but estimates its efficiency to be $7.1\times10^{-28}$ (see Appendix~\ref{sec:lp=m3}). This scattering occurs via transitions involving spin-orbit-coupled modes indicated by the second term on the right-hand side of Eq.~(\ref{eq:OSOI_main}). 
For the aperture angle in this paper, the second term is orders of magnitude smaller than the other terms, resulting in low efficiency. The lack of observation is due to its low efficiency. 
If the aperture angle is increased to, for example, $0.8$, the scattering efficiency is enhanced by 4 orders of magnitude, although it remains relatively small.


Apart from the scattering efficiency, our theoretical framework provides clear implications regarding which $l_p$ modes are, in principle, allowed to be generated. 
As discussed in Sec.~IV, the magneto-optic interaction mediated by spatially uniform magnon modes allows only transitions between spin-orbit-coupled modes that share the same OAM but differ by one in their SAM. 
From this rule and Eq.~(\ref{eq:OSOI_main}), it follows that the generation of optical vortices with $|l_p| = 2$ or $|l_p| \ge 4$ is strictly forbidden. 
One future direction for relaxing this fundamental constraint is to explicitly exploit the discrete rotational symmetry of the crystal, i.e., to utilize rotational umklapp processes~\cite{RA2019}.

The nonreciprocal scattering behavior shown in Figs.~\ref{fig:eff_H} and \ref{fig:eff_invH} results from the breaking of time-reversal symmetry by the external magnetic field. When a quantization axis is defined along the light propagation direction, the magnon SAM $s_m$ changes its sign~\cite{JJ1958} depending on whether the light propagation is parallel or antiparallel to the magnetic ordering direction, which itself is aligned with the external magnetic field. Considering this sign reversal in the context of angular momentum conservation, as described in Eq.~(\ref{eq:conservation}), provides a qualitative explanation for the nonreciprocity of the observed scattering processes.

Since we use magnons in the spatially uniform precession mode, the OAM of the magnon shown in the lower left of Table~\ref{table:AM} does not exist in the first place. On the other hand, magnons in spatially inhomogeneous precession modes, known as magnetic vortex modes~\cite{L1958,PI1959,J1958,MR2004}, possess the OAM. The exploration of the physics governing the BLS induced by magnons in such magnetic vortex modes is an interesting near-term prospect.

Just as the helicity-changing BLS~\cite{RA2019} we previously reported in the same YIG sphere was quickly demonstrated in magnetic thin films~\cite{BY2020,JB2021}, we expect this optical vortex-generating BLS to be demonstrated in various magnetic systems to prove its universality. 
Previous investigations were confined to the paraxial-light regime, where the exchange of angular momentum between light and magnons was discussed only in terms of SAM. 
In contrast, the present work moves beyond this regime by utilizing non-paraxial-light fields, demonstrating that magnons can actively modulate the optical OAM. 
Furthermore, whereas helicity-changing BLS in the paraxial-light regime required in-plane threefold rotational symmetry to satisfy total angular momentum conservation law~\cite{RA2019,BY2020,JB2021}, the inclusion of the optical OAM term in the conservation law---as shown in Eq.~(\ref{eq:conservation})---relaxes these requirements on the material side. 
Consequently, magnon-induced optical vortex generation enables magneto-optical studies for a broad range of magnetic materials regardless of their crystal symmetry.

Finally, we discuss the possibility of dynamically modulating the OAM of propagating light using the phenomenon presented in this paper. OAM represents an untapped degree of freedom in optical communications~\cite{AH2015,J2016,AK2021,JJ2022}. The reasons for this are the lack of physics enabling high-speed OAM modulation and the absence of fibers capable of preserving OAM. Magnons with linewidths exceeding megahertz can change the BLS they mediate at equivalent speeds. Magnon-based optical vortex generation could be controlled with response speeds exceeding megahertz, offering a new approach to addressing the former challenge.

In summary, focused spin-orbit-coupled light under magnetic ordering proved to exhibit rich physics arising from the time-reversal symmetry breaking, which may open a new panorama for topological photonics~\cite{MJ2014,LJ2014}, optomagnonics~\cite{BS2022}, and chiral quantum optics~\cite{PS2017}. Further diverse scattering phenomena will be revealed by examining the interaction between magnon modes with spatial structure and the spin-orbit-coupled light. Conversely, the intricate scattering processes involving the spatially structured magnons and photons in magnetic materials could be explored by utilizing the composite physics described in this work.

\begin{table}[t]
  \begin{center}
    \caption{Change in angular momenta} 
    \begin{tabular}{|c||c|c|} \hline
       & \begin{tabular}{c}Magnon\\ in uniform mode \end{tabular}& Photon   \\ \hline \hline 
      Spin & $|\Delta s_m|=1$ &$|\Delta s_p|=0,2$ \\ \hline 
      Orbital & - & $|\Delta l_p|=0,1,2,\ldots$ \\ \hline
    \end{tabular}
  \label{table:AM}
  \end{center}
\end{table}

\acknowledgments
 We thank K.~Usami, Y.~Nakata, and R.~Inoue for useful discussions, and Y.~Nakamura for lending us YIG sphere. 
 This work was supported by Japan Science and Technology Agency (JST) PRESTO Grant No. JPMJPR200A (R.H.); JST ASPIRE Grant No. JPMJAP2409 (T.O.); Japan Society for the Promotion of Science (JSPS) KAKENHI Grants No. JP22K14589 (R.H.), No. JP24H02233 (Y.S.), and No. JP25K00937 (R.H.); HONMAMON co-creative research program between HORIBA and Kyoto University (R.H.); and the Cooperative Research Project of Research Institute of Electrical Communication (RIEC), Tohoku University (T.O.).
 
The authors declare that they have no competing interests.

\section*{Data Availability}
The data described in the manuscript is available in the Zenodo database~\cite{RA2026}.

\appendix
\section{Methods}
\subsection{YIG sphere}
We purchased the YIG spherical crystal from Ferrisphere Inc. The sample is polished to the surface roughness below $50\,\rm{nm}$ and glued to an aluminum-oxide rod at the factory.

\subsection{Optical measurements} \label{sec:methods_opto}
To investigate the scattering of the optical Gaussian beam to the optical vortex beam and its scattering rule, we distinguish the SAM (helicity), OAM, and frequencies of the input and scattered beams under conditions where magnons in uniform mode are continuously driven at resonance. As shown in Fig.~\ref{fig:setup}(b), a CW laser light with a wavelength of $1550\,\rm{nm}$ (angular frequency of $\Omega_C$) is split into two paths by a fiber splitter. In the upper path, the Gaussian laser beam output from an SM fiber is sent through the center of the YIG sphere, parallel to the $\langle100\rangle$ axis of the YIG monocrystal, as shown in Fig.~\ref{fig:setup}(a). Since magnons in the uniform mode have no wave number, scattered photons propagate coaxially with transmitted photons. 
By a QWP–PBS pair before and after the YIG sphere, either the left or right circularly polarized light corresponding to the negative or positive SAM of the photon can be selected as the input and output.

We determine the frequencies of the input and scattered light using a heterodyne technique. The light in the lower path in Fig.~\ref{fig:setup}(b) is frequency shifted by $\Omega_A/2\pi=110\,\rm{MHz}$ from $\Omega_C$ by an AOM and is used as an LO to identify the frequency of scattered light. As shown schematically in Figs.~\ref{fig:setup}(b) and \ref{fig:setup}(c), the scattered light from the upper path interferes with the LO light from the lower path after the second fiber splitter so that the resultant beat signals originating from Stokes scattering (red sideband) and anti-Stokes scattering (blue sideband) appear at different angular frequencies, $\omega_R$ and $\omega_B$, respectively. These beat signals are detected by a high-speed photodetector, amplified by a microwave amplifier, and fed into a spectrum analyzer.

We determine the OAM of the scattered photons, $l_s \hbar$, using an original detection system shown in Fig.~\ref{fig:setup}(a), which consists of an SLM with a computer-controlled liquid crystal pattern and an SM fiber. This system works on the following principle. First, under appropriate alignment, the SLM converts scattered photons with $\mathrm{OAM}=l_s \hbar$ into reflected photons with $\mathrm{OAM}=l_r \hbar$. There is a relationship between $l_s$ and $l_r$ of $l_s-l_r=C$, and this $C$ is an integer value that we can arbitrarily set through the liquid crystal pattern of the SLM. Furthermore, when the reflected photons are input into the SM fiber via an appropriate lens, only the Gaussian mode ($l_r=0$) can pass through, creating a situation where other modes are reflected. Note that independent experiments confirm that the extinction ratio of this method is more than $20\,\rm{dB}$. Under such circumstances, only the scattered light, which initially has $l_s=C$, can reach the HPD. Figure~\ref{fig:setup}(d) shows the relationship between the liquid crystal pattern used in this experiment and $l_s$ determined from it.

Furthermore, this study evaluates the scattering efficiency into the LG modes with $p=0$. Generally, controlling only the optical phase via an SLM cannot completely convert any $p \neq 0$ mode into the fundamental Gaussian mode. However, in our experimental configuration, this characteristic works to the advantage of selective mode detection. Even if the SLM flattens the phase front, the concentric amplitude profile inherent to $p \neq 0$ modes remains; as a result, the overlap integral with the mode of the SM fiber is minimized, and their coupling is significantly suppressed. Because the SM fiber itself functions as a strict spatial filter, only the scattered modes with $p=0$ can be selectively coupled, effectively eliminating potential contamination from higher-order radial scattering modes.

\section{Scattering efficiencies when external magnetic field is reversed}
Figures~\ref{fig:eff_invH}(a)-\ref{fig:eff_invH}(d) show the results when the direction of the magnetic field is reversed from that in Figs.~\ref{fig:eff_H}(a)-\ref{fig:eff_H}(d). Note that the sign of $\Delta s_m$ associated with Stokes or anti-Stokes scattering is reversed compared to that in Figs.~\ref{fig:eff_H}(a)-\ref{fig:eff_H}(d) due to the reversal of the external magnetic field.

\begin{figure*}[t]
\begin{center}
\includegraphics[width=18.0cm,angle=0]{BarPlots_invH_Kittel_100axis_W180_rev20_W1800_PRstyle.pdf}
\caption{
(a)-(d) Scattering efficiencies of the Stokes sidebands (red bars) and the anti-Stokes sidebands (blue bars) for four distinct optical polarization sets under conditions where the external magnetic field direction is reversed compared to Figs.~3(a)-3(d). The height of the color bar shows the mean scattering efficiency, and the difference between the top of the black wireframe and the bar represents a standard deviation estimated from measurements repeated six times. The gray arrows indicate scattering that satisfies the conservation of total angular momentum.
}
\label{fig:eff_invH}
\end{center}
\end{figure*}

\section{Scattering efficiencies with $|\Delta l_p|\le3$} \label{sec:lp=3}
\begin{figure}[t]
\begin{center}
\includegraphics[width=7.5cm,angle=0]{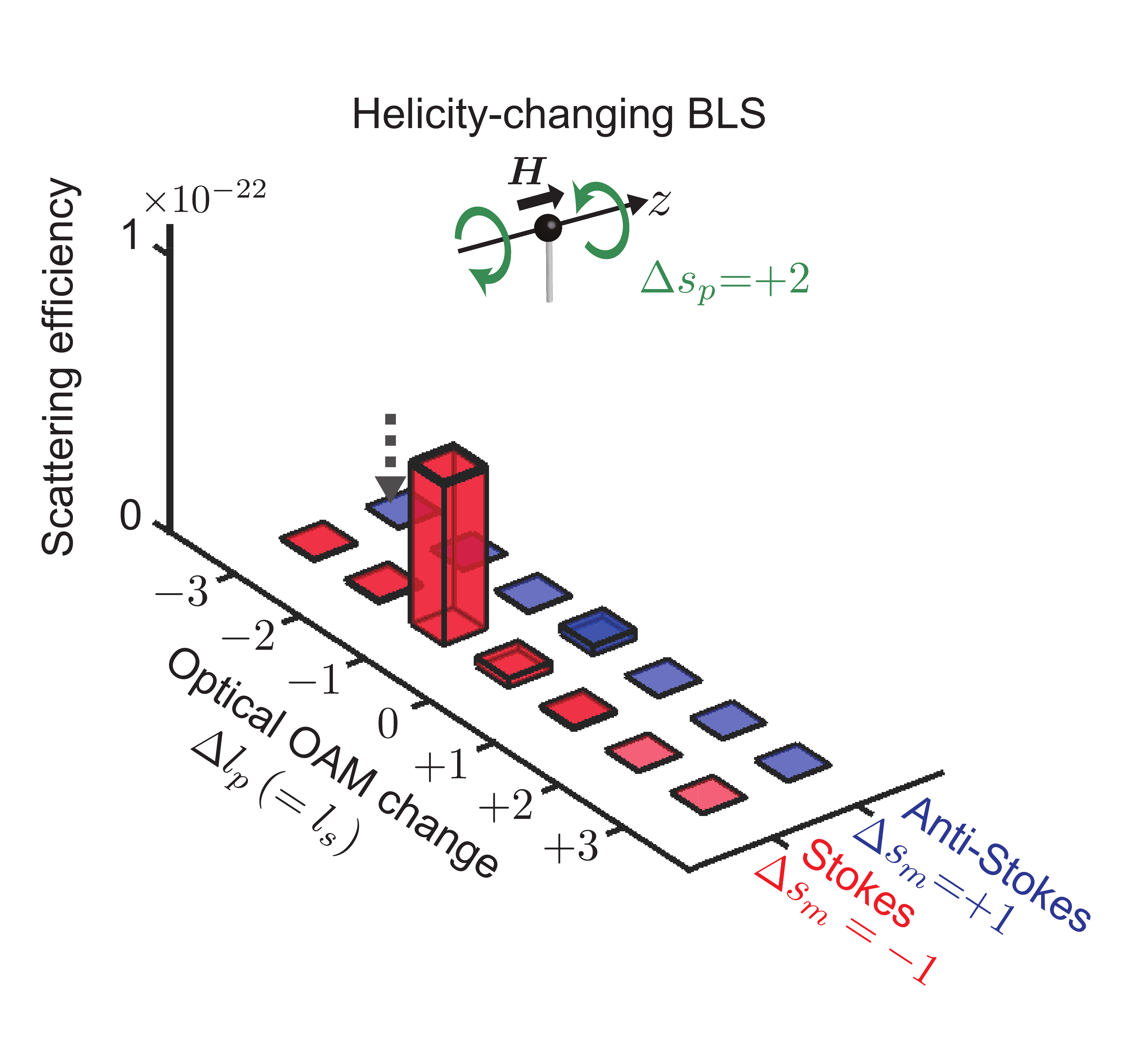}
\caption{
Scattering efficiencies of the Stokes sidebands (red bars) and the anti-Stokes sidebands (blue bars) for $|\Delta l_p|\le3$.
}
\label{fig:eff_H_LG03}
\end{center}
\end{figure}

Figure~\ref{fig:eff_H_LG03} shows the results of Fig.~\ref{fig:eff_H}(c) with the addition of the results for $|\Delta l_p |=3$. The gray dashed line in Fig.~\ref{fig:eff_H_LG03} shows the scattering with the combination $(\Delta s_m, \Delta s_p, \Delta l_p)=(+1, +2, -3)$ that satisfies the conservation of total angular momentum, although it is not observed.

\section{Details of theory} \label{sec:details_theory}
The electric field of the nonparaxial light focused by a convex lens has been extensively studied. Now, we focus on a monochromatic circularly polarized optical vortex propagating along the $z$ axis as an input paraxial light. The input light is transformed into spin-orbit-coupled (nonparaxial) light through refraction at the spherical surface, as shown in Fig.~\ref{fig:concept}(b). In investigating the angular momentum transfer during BLS, it is instrumental to invoke the spherical basis. The electric field in the Cartesian basis is given by
\begin{equation}
\bm{E}=E_x\bm{e}_x+E_y\bm{e}_y+E_z\bm{e}_z, \label{eq:E_cart}
\end{equation}
while in the spherical basis it is
\begin{equation}
    \bm{E}=E_R\bm{e}_R+E_0\bm{e}_0+E_L\bm{e}_L, \label{eq:E_sphe}
\end{equation}
where the spherical basis $\{\bm{e}_R, \bm{e}_0, \bm{e}_L\}$ is related to the Cartesian basis $\{\bm{e}_x, \bm{e}_y, \bm{e}_z\}$ as
\begin{equation}
    \left[
\begin{array}{c}
\bm{e}_R\\
\bm{e}_0\\
\bm{e}_L
\end{array}
\right]
=
 \left[
\begin{array}{c}
\bm{e}_{-1}\\
\bm{e}_0\\
\bm{e}_{+1}
\end{array}
\right]
=
 \left[
\begin{array}{c}
-\frac{1}{\sqrt{2}}(\bm{e}_x+i\bm{e}_y)\\
\bm{e}_z\\
\frac{1}{\sqrt{2}}(\bm{e}_x-i\bm{e}_y)
\end{array}
\right]. \label{eq:E_cart_sphe}
\end{equation}
Here, along the reference axis ($z$ axis), $E_R$ and $E_L$ are the right- and left-circularly polarized components of the electric field, respectively, while $E_0$ is the longitudinal component. In the spherical basis, the relationship between the far-field-electric-field mode and the spin-orbit-coupled-light modes within the sphere is expressed as~\cite{KE2011} 
\begin{widetext}
\begin{equation}
\bm{E}=Ee^{il_p\phi}\bm{e}_{s_p} =\sqrt{\cos\theta}\left(aEe^{il_p\phi}\bm{e}_{s_p}-bEe^{i(l_p+2s_p)\phi}\bm{e}_{-s_p}-\sqrt{2ab}Ee^{i(l_p+s_p)\phi}\bm{e}_{0}\right),   \label{eq:OptSOI} 
\end{equation}
\end{widetext}
where $\{s_p\}\in\{+1, 0, -1\},\,\{l_p\}\in\mathbb{Z},\,a=\cos^2(\theta/2),\,b=\sin^2(\theta/2),\,E$ is the amplitude, $\phi$ is the azimuthal angle, and $\theta$ is defined as the aperture angle shown in Fig.~\ref{fig:concept}(b). 
Physically, even as the wave vectors of paraxial light undergo angular spreading upon refraction with the aperture angle $\theta$, each local ray preserves the wave transversality condition ($\bm{k} \cdot \bm{E} = 0$). Representing these fields in the laboratory spherical basis $\bm{e}_{s_p}$ [Eq.~(\ref{eq:E_cart_sphe})] via a geometric basis transformation gives the coefficients on the right-hand side of Eq.~(\ref{eq:OptSOI}); these serve as mode-coupling coefficients that guarantee energy conservation. Furthermore, the total angular momentum, $l_p + s_p$, is conserved between the paraxial far-field mode on the left-hand side and each spin-orbit-coupled-light mode on the right-hand side. In the absence of absorption or loss, refraction at the spherical boundary is time-reversal symmetric; thus, Eq.~(\ref{eq:OptSOI}) governs both the forward conversion from the paraxial mode into spin-orbit-coupled-light modes and its reverse process. In this reverse process, as light diverging within the aperture angle $\theta$ refracts back into a collimated paraxial beam, the internal spin-orbit-coupled-light modes on the right-hand side are mapped back onto the far-field paraxial mode on the left-hand side.
To simplify future discussions, we rewrite Eq.~(\ref{eq:OptSOI}) using the Laguerre-Gauss mode ${\rm{LG}}_{l_p,s_p}$ as
\begin{widetext}
\begin{equation}
{\rm{LG}}_{l_p,s_p}=\sqrt{\cos\theta}\left(a\times{\rm{LG}}_{l_p,s_p}^{\rm{SO}}-b\times{\rm{LG}}_{l_p+2s_p,-s_p}^{\rm{SO}}-\sqrt{2ab}\times{\rm{LG}}_{l_p+s_p,0}^{\rm{SO}}\right). \label{eq:LG_OptSOI}
\end{equation}
\end{widetext}
Here, we denote optical modes as ${\rm{LG}}_{l_p,s_p}$ with OAM in the first subscript and SAM in the second subscript.

The classical interaction Hamiltonian representing the BLS is given by~\cite{RA2019}
\begin{equation}
    H_{\rm{int}}(\tau)=\frac{\epsilon_0}{2}\int^{\tau+\delta}_{\tau}\mathcal{E}Ac'dt, \label{eq:Hint}
\end{equation}
where $\epsilon_0$ is the permittivity of free space, $A$ is the cross section of the light beam, and $\delta=l/c'$ is the interaction time, with $c'$ being the speed of light in the material and $l$ being the interaction length. Here,
\begin{equation}
    \mathcal{E}=
        \left[
\begin{array}{ccc}
E^{'*}_{x}  & E^{'*}_{y} & E^{'*}_{z} \\
\end{array}
\right]
    \left[
\begin{array}{ccc}
\epsilon_{xx}  & \epsilon_{xy} & \epsilon_{xz} \\
\epsilon_{yx} & \epsilon_{yy} & \epsilon_{yz} \\
\epsilon_{zx} & \epsilon_{zy} & \epsilon_{zz} \\
\end{array}
\right]
        \left[
\begin{array}{ccc}
E_x \\
E_y\\
E_z \label{eq:EneDens_cart}
\end{array}
\right]
\end{equation}
is the Hamiltonian density, with $\epsilon_{ij}$ being the $ij$ component of the second rank dielectric tensor, and $E_i$ and $E^{'*}_{i}$ being the $i$ components of the incident and scattered electric field, respectively. Phenomenologically, we can understand the BLS by considering the tensor $\epsilon_{ij}$ in Eq.~(\ref{eq:EneDens_cart}) as a function of the magnetization. With the spherical basis, the Hamiltonian density in Eq.~(\ref{eq:EneDens_cart}) can be rewritten as
\begin{equation}
    \mathcal{E}=
        \left[
\begin{array}{ccc}
E^{'*}_{R}  & E^{'*}_{0} & E^{'*}_{L} \\
\end{array}
\right]
    \left[
\begin{array}{ccc}
\epsilon_{RR}  & \epsilon_{R0} & \epsilon_{RL} \\
\epsilon_{0R} & \epsilon_{00} & \epsilon_{0L} \\
\epsilon_{LR} & \epsilon_{L0} & \epsilon_{LL} \\
\end{array}
\right]
        \left[
\begin{array}{ccc}
E_R \\
E_0\\
E_L 
\end{array}
\right]. \label{eq:EneDens_sphe}
\end{equation}

The dielectric tensor $\epsilon_{ij}$, describing the strength of the coupling between light and magnetization, is endowed with crystal symmetry and the Onsager relation~\cite{BD1967}. Furthermore, in situations where optical absorption is negligible, as in this paper’s case, the additional requirement is that the dielectric tensor is Hermitian. 
The derivation of the specific form of the dielectric tensor for cubic crystals such as YIG is summarized in Sec.~S1 of the Supplemental Material to our paper~\cite{RA2019}. Based on that, when an external magnetic field of enough strength to magnetize the crystal is applied parallel to the $\langle100\rangle$ crystal axis and $z$ axis, the dielectric tensor with the spherical basis can be expressed as
\begin{widetext}
\begin{equation}
\epsilon =\frac{1}{\sqrt{2}}    
    \left[
\begin{array}{ccc}
0  & fm^{-}(t)-2G_{44}m_zm^{-}(t) & 0 \\
-fm^{+}(t)+2G_{44}m_zm^{+}(t) & 0 & fm^{-}(t)+2G_{44}m_z m^{-}(t) \\
0 & -fm^{+}(t)-2G_{44}m_z m^{+}(t) & 0 \\
\end{array}
\right], \label{eq:eps_cart}
\end{equation}
\end{widetext}
where $f$ is a parameter representing the magnitude of the Faraday effect and $G_{44}$ is one of three parameters representing the magnitude of the Cotton-Mouton effect. Following the experiment in this paper, we assume that magnons in uniform mode, called Kittel mode, are excited in a ferromagnetic crystal. The resultant time-varying magnetization vector in the Cartesian basis is represented by
\begin{equation}
    \bm{m}(t)=
            \left[
\begin{array}{ccc}
m_x(t)\\ 
m_y(t)\\ 
m_z
\end{array}
\right]. \label{eq:vec_mag}
\end{equation}
The mean magnetization $m_z$ along the $z$ axis is much larger than the magnetization in the plane perpendicular to the $z$ axis; therefore, $m_z$ can be considered constant and identical to the saturation magnetization $m_s$. $m^+(t)=m_x(t)+im_y (t)$ and $m^-(t)=m_x (t)-im_y(t)$ are the normal modes of the transverse magnetizations. Note that we display only the terms associated with a single magnon excitation in the quantum mechanical interpretation in Eq.~(\ref{eq:eps_cart}). Quantum mechanically, $m^+$ ($m^-$) annihilates (creates) a magnon and increases (reduces) the $z$ component of the magnetization, as we explain later. Equation~(\ref{eq:eps_cart}) indicates that the probabilities of BLS between $R$ and $z$-polarized light and between $L$ and $z$-polarized light are nonzero, and the terms with $m^+$ ($m^-$) represent the anti-Stokes scattering (Stokes scattering). 

The uniform-mode magnon scatters spin-orbit-coupled light within the YIG sphere. Since the magnon has a wave number close to zero, the input and scattered light beams retain all beam parameters except polarization and frequency. The nonzero terms of the dielectric tensor in Eq.~(\ref{eq:eps_cart}) contribute to scattering the input spin-orbit-coupled-light mode in Eq.~(\ref{eq:LG_OptSOI}) into another spin-orbit-coupled-light mode with different polarization but the same OAM. The scattered light undergoes refraction upon exiting the sphere and is recovered into far-field-light (paraxial-light) mode, resulting in scattered far-field (paraxial) light with a different state from the input far-field light.

Now, let us examine the above scattering process specifically using the observed Stokes scattering with gray arrows in Fig.~\ref{fig:eff_H}(a), as shown in Fig.~\ref{fig:concept}(b). Following Eq. (\ref{eq:LG_OptSOI}), writing down the correspondence between the far-field mode and the spin-orbit-coupled-light modes inside the sphere for the input-Gaussian-paraxial-light mode and the scattered-paraxial-vortex-light mode yields
\begin{widetext}
\begin{equation}
    {\rm{LG}}_{0,+1}=\sqrt{\cos\theta}\left(a\times{\rm{LG}}_{0,+1}^{\rm{SO}}-b\times{\rm{LG}}_{+2,-1}^{\rm{SO}}-\sqrt{2ab}\times{\rm{LG}}_{+1,0}^{\rm{SO}}\right) \label{eq:LG01_OptSOI}
\end{equation}
and
\begin{equation}
    \sqrt{\cos\theta}\left(a\times{\rm{LG}'}_{+1,+1}^{\rm{SO}}-b\times{\rm{LG}'}_{+3,-1}^{\rm{SO}}-\sqrt{2ab}\times{\rm{LG}'}_{+2,0}^{\rm{SO}}\right)={\rm{LG}'}_{+1,+1}. \label{eq:LG11_OptSOI}
\end{equation}
\end{widetext}
The prime symbol in Eq.~(\ref{eq:LG11_OptSOI}) is assigned to the scattered light. The Stokes scatterings between spin-orbit-coupled-light modes described by Eqs.~(\ref{eq:Hint}) and (\ref{eq:eps_cart}) occur between ${\rm{LG}}_{+1,0}^{\rm{SO}}$ and ${\rm{LG}'}_{+1,+1}^{\rm{SO}}$, and between ${\rm{LG}}_{+2,-1}^{\rm{SO}}$ and ${\rm{LG}'}_{+2,0}^{\rm{SO}}$, which share the same OAM. Here, we assume that spatial phase matching between the input and the scattered light modes holds when magnons are spatially uniform. The consistency between experimental and theoretical scattering efficiencies finally validates this assumption.

To calculate the scattering efficiency, which refers to the probability that one magnon scatters one incident photon, we quantize the classical interaction Hamiltonian in Eq. (\ref{eq:Hint}). As preparation for this, we quantize the magnetization and the electric field. First, let the magnetic moment operators be $\{\hat{M}_x,\hat{M}_y,\hat{M}_z\}$, which satisfy the standard commutation relation
\begin{equation}
            \left[
\begin{array}{ccc}
\hat{M}_i,& \hat{M}_j 
\end{array}
\right]
=
i\epsilon_{ijk}\mu_B\hat{M}_k, \,\{i,j,k\}=\{x,y,z\}, \label{eq:commutation_M}
\end{equation}
where $\hbar=1$, $\epsilon_{ijk}$ is Levi-Civita symbol, and $\mu_B$ is the Bohr magneton. With the linearized version of the Holstein-Primakoff transformation, the magnetic moment operators are represented by bosonic magnon operators $\hat{b}^\dagger$ and $\hat{b}$:
\begin{eqnarray}
\hat{M}_x&=& -\frac{\mu_B}{2}\sqrt{N}(\hat{b}+\hat{b}^\dagger),\\
\hat{M}_y&=& -\frac{\mu_B}{2i}\sqrt{N}(\hat{b}-\hat{b}^\dagger),\\
\hat{M}_z&=& -\frac{\mu_B}{2}N, \label{eq:M_b_relation} 
\end{eqnarray}
which are valid when $N\gg1$, with $N$ being the total spin number participated in the interaction. Note here that the operators $\hat{b}^\dagger$ and $\hat{b}$ are dimensionless and fulfill the commutation relation $[\hat{b},\hat{b}^\dagger]=1$. This commutation relation implies that these operators represent zero-dimensional modes such as modes within a resonator. The magnetization operators can then be given by
\begin{eqnarray}
\hat{m}_x&=& \frac{\hat{M}_x}{V_s}=-\frac{\mu_B}{2V_s}\sqrt{N}(\hat{b}+\hat{b}^\dagger),\\
\hat{m}_y&=& \frac{\hat{M}_y}{V_s}=-\frac{\mu_B}{2iV_s}\sqrt{N}(\hat{b}-\hat{b}^\dagger),\\
\hat{m}_z&=& \frac{\hat{M}_z}{V_s}=-\frac{\mu_B}{2V_s}N, \label{eq:m_b_relation} 
\end{eqnarray}
where $V_s$ is the sample volume. The magnetization operators satisfy the following commutation relation:
\begin{equation}
            \left[
\begin{array}{ccc}
\hat{m}_i,& \hat{m}_j 
\end{array}
\right]
=
i\epsilon_{ijk}\frac{\mu_B}{V_s}\hat{m}_k, \,\{i,j,k\}=\{x,y,z\}. \label{eq:commutation_m}
\end{equation}
The magnetization ladder operators, $\hat{m}^+$ and $\hat{m}^-$, are accordingly defined as
\begin{eqnarray}
\hat{m}^+ &=& \hat{m}_x +i\hat{m}_y=-\frac{\mu_B \sqrt{n}}{\sqrt{V_s}}\hat{b}, \label{eq:mp_cart2sphe}\\
\hat{m}^- &=& \hat{m}_x -i\hat{m}_y=-\frac{\mu_B \sqrt{n}}{\sqrt{V_s}}\hat{b}^\dagger, \label{eq:mm_cart2sphe}
\end{eqnarray}
where $n=N/V_s$ is the spin density. These equations link the creation and annihilation of magnons with $\hat{m}^+$ and $\hat{m}^-$.

Next, we summarize the continuum description of quantized electromagnetic fields. The derivation is well documented in Refs.~\cite{KR1990,MT1995}. An electric field propagating in one direction with a single frequency and polarization $\sigma$ ($\{\sigma\}=\{R,0,L\}$) is represented by bosonic photon operators $\hat{a}_\sigma^\dagger$ and $\hat{a}_\sigma$:
\begin{equation}
    \hat{E}_\sigma=\sqrt{\frac{\hbar \Omega}{2\epsilon_0 Ac}}\left(\hat{a}_\sigma(t)+\hat{a}^\dagger(t)\right), \label{eq:E_qunatum}
\end{equation}
where $\Omega$ is the angular frequency of the electric field, $A$ is the transverse cross section, and $c$ is the speed of light in vacuum. Note here that the operators $\hat{a}_\sigma^\dagger(t)$ and $\hat{a}_\sigma(t)$ are continuous as a function of $t$ and they have units of square root times. The operators fulfill the commutation relation $[\hat{a}_\sigma(t),\hat{a}^{\dagger}_{\sigma'}(t')]=\delta(t-t')\delta_{\sigma\sigma'}$. The physical interpretation is that these operators represent one-dimensional modes, that is, itinerant modes and $\hat{a}_\sigma^\dagger(t)\hat{a}_\sigma(t)$ is the flux of photons at time $t$ at a position.

Finally, we convert the electric fields and the magnetization in Eqs.~(\ref{eq:EneDens_sphe}) and (\ref{eq:eps_cart}) into quantum mechanical operators by the following simple prescription:
\begin{equation}
            \left[
\begin{array}{cccc}
E_\sigma\\ 
E_\sigma^*\\ 
m^+\\
m^-
\end{array}
\right]
\rightarrow
            \left[
\begin{array}{cccc}
\hat{E}_\sigma\\ 
\hat{E}_\sigma^\dagger\\ 
\hat{m}^+\\
\hat{m}^-
\end{array}
\right].
\label{eq:prescription}
\end{equation}
Then, using Eqs.~(\ref{eq:mp_cart2sphe}), (\ref{eq:mm_cart2sphe}), and (\ref{eq:E_qunatum}), the Hamiltonian in Eq.~(\ref{eq:Hint}) is rewritten as
\begin{widetext}
    \begin{equation}
        \hat{H}_{\rm{int}}(\tau)=\frac{\epsilon_0}{2}\int^{\tau+\delta}_{\tau}\hat{E}_{\sigma'}^{'\dagger}\hat{\epsilon}_{\sigma'\sigma}\hat{E}_\sigma A c' dt +{\rm{H.c.}} 
        =\frac{\hbar\sqrt{\Omega_0 \Omega' }l}{4c}\left(\hat{a}'_{\sigma'}(\tau)+\hat{a}^{'\dagger}_{\sigma'}(\tau)\right)\epsilon_{\sigma'\sigma}\left(\hat{a}_\sigma(\tau)+\hat{a}_\sigma^\dagger(\tau)\right)+\rm{H.c.}, \label{eq:Hint_quantum}
    \end{equation}
\end{widetext}
where in the last step we use the fact that the duration of the interaction is shorter than any other time scales of the dynamics and $\int^{\tau}_{0}c'dt=l$. Furthermore, here, $\Omega_0$ ($\Omega'$) is the angular frequency of the incident (scattered) light and we use Einstein notation. The dielectric tensor in Eq.~(\ref{eq:Hint_quantum}) is also quantized as
\begin{widetext}
\begin{equation}
\hat{\epsilon}_{\sigma'\sigma} =\frac{-\mu_B\sqrt{n}}{\sqrt{2V_s}}    
    \left[
\begin{array}{ccc}
0  & \left(f+2G_{44}\left(\frac{\mu_B}{2V_s}N\right)\right)\hat{b}^\dagger(t) & 0 \\
\left(-f-2G_{44}\left(\frac{\mu_B}{2V_s}N\right)\right)\hat{b}(t) & 0 & \left(f-2G_{44}\left(\frac{\mu_B}{2V_s}N\right)\right)\hat{b}^\dagger(t) \\
0 & \left(-f+2G_{44}\left(\frac{\mu_B}{2V_s}N\right)\right)\hat{b}(t)  & 0 \\
\end{array}
\right]. \label{eq:eps_quantum}
\end{equation}
\end{widetext}
With the rotating wave approximation, if the angular frequency $\Omega'$ of the scattered light $\hat{a}'_{\sigma'}$ is $\Omega'=\Omega_0+\omega_m$, with $\Omega_0$ being the frequency of the incident light and $\omega_m$ being that of the magnon, the approximated Hamiltonian becomes
\begin{equation}
    \hat{H}_{\rm{int}}\simeq \frac{-\hbar \sqrt{\Omega_0\Omega'}l\mu_B\sqrt{n}}{4\sqrt{2}\sqrt{V_s}c} \hat{\xi}_{\sigma'\sigma}(\hat{a}^{'\dagger}_{\sigma'}\hat{b}\hat{a}_\sigma+\hat{a}^{'}_{\sigma'}\hat{b}^\dagger\hat{a}^\dagger_{\sigma}). \label{eq:Hint_quantum_1}
\end{equation}
With the rotating wave approximation, if the frequency $\Omega'$ is $\Omega'=\Omega_0-\omega_m$, the approximated Hamiltonian becomes
\begin{equation}
    \hat{H}_{\rm{int}}\simeq \frac{-\hbar \sqrt{\Omega_0\Omega'}l\mu_B\sqrt{n}}{4\sqrt{2}\sqrt{V_s}c} \hat{\xi}_{\sigma'\sigma}(\hat{a}^{'}_{\sigma'}\hat{b}\hat{a}_\sigma^\dagger+\hat{a}^{'\dagger}_{\sigma'}\hat{b}^\dagger\hat{a}_{\sigma}), \label{eq:Hint_quantum_2}
\end{equation}
where
\begin{widetext}
\begin{equation}
\hat{\xi}_{\sigma'\sigma} =    
    \left[
\begin{array}{ccc}
0  & \left(f+2G_{44}\left(\frac{\mu_B}{2V_s}N\right)\right) & 0 \\
\left(-f-2G_{44}\left(\frac{\mu_B}{2V_s}N\right)\right) & 0 & \left(f-2G_{44}\left(\frac{\mu_B}{2V_s}N\right)\right) \\
0 & \left(-f+2G_{44}\left(\frac{\mu_B}{2V_s}N\right)\right)  & 0 \\
\end{array}
\right]. \label{eq:ene_density_quantum}
\end{equation}
\end{widetext}
Equations~(\ref{eq:Hint_quantum_1}) and (\ref{eq:Hint_quantum_2}) imply the conservation laws of energy and SAM between light and magnons during BLS in the system with rotational symmetry. Given that initially the incident light is circularly polarized, the operator $\hat{a}_\sigma$ ($\hat{a}^\dagger_\sigma$) can be treated as the classical amplitude $\beta_\sigma$ ($\beta^*_\sigma$), where we find $|\beta_\sigma|^2=\frac{P_\sigma}{\hbar\Omega_0}$; i.e., $|\beta_\sigma|^2$ is the total incident photon flux. The Hamiltonian in Eq.~(\ref{eq:Hint_quantum_1}) becomes the beam splitter type:
\begin{equation}
    \hat{H}_{\sigma'\sigma}=-\hbar\sqrt{\zeta_{\sigma'\sigma}}(\hat{a}^{'\dagger}_{\sigma'}\hat{b}+\hat{a}'_{\sigma'}\hat{b}^\dagger), \label{eq:Hint_BS}
\end{equation}
while the Hamiltonian in Eq.~(\ref{eq:Hint_quantum_2}) becomes the parametric amplifier type:
\begin{equation}
    \hat{H}_{\sigma'\sigma}=\hbar\sqrt{\zeta_{\sigma'\sigma}}(\hat{a}^{'}_{\sigma'}\hat{b}+\hat{a}^{'\dagger}_{\sigma'}\hat{b}^\dagger), \label{eq:Hint_PA}
\end{equation}
where $\zeta_{\sigma'\sigma}$ is
\begin{widetext}
\begin{eqnarray}
\zeta_{\sigma'\sigma} 
&=&
    \left[
\begin{array}{ccc}
\zeta_{RR} & \zeta_{R0}  & \zeta_{RL}  \\
\zeta_{0R}  & \zeta_{00} & \zeta_{0L} \\
\zeta_{LR} & \zeta_{L0}  & \zeta_{LL} \\
\end{array}
\right]\notag\\
&=&
\frac{\Omega_0 \Omega' l^2\mu_B^2n}{32V_s c^2}\left(\frac{P_\sigma}{\hbar\Omega_0}\right)\left[
\begin{array}{ccc}
0  & \left(f+2G_{44}\left(\frac{\mu_B}{2V_s}N\right)\right)^2 & 0 \\
\left(-f-2G_{44}\left(\frac{\mu_B}{2V_s}N\right)\right)^2 & 0 & \left(f-2G_{44}\left(\frac{\mu_B}{2V_s}N\right)\right)^2 \\
0 & \left(-f+2G_{44}\left(\frac{\mu_B}{2V_s}N\right)\right)^2  & 0 \\
\end{array}
\right]
. \label{eq:zeta}
\end{eqnarray}
\end{widetext}
Note that the $\zeta_{\sigma'\sigma}$ has the dimension of angular frequency.

The input-output theory in quantum mechanics defines the boundary condition between $\hat{a}'_{\sigma'}$ and $\hat{b}$, which are coupled by the Hamiltonian equation~(\ref{eq:Hint_BS}) or (\ref{eq:Hint_PA}), as~\cite{AM2010} $\hat{a}'_{\sigma'}=\sqrt{\zeta_{\sigma'\sigma}}\hat{b}$. This corresponds to radiation from the uniform magnon mode $\hat{b}$ into the scattered photon mode $\hat{a}'_{\sigma'}$. From the boundary condition, the relationship between the scattered photon flux and the number of magnons is given by
\begin{equation}
    \langle \hat{a}^{'\dagger}_{\sigma'}\hat{a}^{'}_{\sigma'}\rangle=\frac{\zeta_{\sigma'\sigma}}{2\pi}\langle\hat{b}^\dagger\hat{b}\rangle. \label{eq:relation_aa_bb}
\end{equation}
Here, $\langle\cdots\rangle$ refers to the expected value of the coherent state corresponding to a specific mode for each experiment. Furthermore, from Eqs.~(\ref{eq:zeta}) and (\ref{eq:relation_aa_bb}), the probability $\eta_{\sigma'\sigma}$ that single magnon scatters an input spin-orbit-coupled-photon mode into another spin-orbit-coupled-photon mode is described as
\begin{equation}
    \eta_{\sigma'\sigma}=\frac{\langle\hat{a}^{'\dagger}_{\sigma'}\hat{a}'_{\sigma'}\rangle}{\langle\hat{b}^\dagger\hat{b}\rangle\langle\frac{P_\sigma}{\hbar \Omega_0}\rangle}=\frac{\zeta_{\sigma'\sigma}}{2\pi}\left(\frac{P_\sigma}{\hbar\Omega_0}\right)^{-1}. \label{eq:probability}
\end{equation}
Note that $\eta_{\sigma'\sigma}$ is dimensionless.

The scattering efficiencies between the input and scattered far fields are determined by the product of the energy partitioning ratio between the far field and spin-orbit-coupled light during two refractions at the sphere surface, which depends on aperture angle, and the BLS efficiency between spin-orbit-coupled light within the YIG sphere.

\subsection{$L_i\rightarrow L_o$ configuration} \label{sec:details_theory_Li2Lo}
Let us calculate the overall scattering efficiency of the optical-vortex-Stokes scattering shown as the significant red bar with the gray arrow in Fig.~\ref{fig:eff_H}(a). The transition process from the input-far-field mode ${\rm{LG}}_{0,+1}$ to the scattered-optical-vortex-far-field mode ${\rm{LG}}'_{+1,+1}$ involves two pathways via BLS within the YIG sphere: one from ${\rm{LG}}_{+1,0}^{\rm{SO}}$ to ${\rm{LG}'}^{\rm{SO}}_{+1,+1}$ and another one from ${\rm{LG}}_{+2,-1}^{\rm{SO}}$ to ${\rm{LG}'}_{+2,0}^{\rm{SO}}$, as shown in Fig.~\ref{fig:concept}(b). For the former path, the initial spherical refraction first converts a $(\sqrt{2ab\times\cos \theta})^2$ of the far-field mode ${\rm{LG}}_{0,+1}$ into spin-orbit-coupled-light mode ${\rm{LG}}_{+1,0}^{\rm{SO}}$ in the dimension of power. This then undergoes BLS with an efficiency $\eta_{L0}$ into spin-orbit-coupled-light mode ${\rm{LG}'}_{+1,+1}^{\rm{SO}}$, and finally, spherical refraction converts a $(a\sqrt{\cos \theta})^2$ of the ${\rm{LG}'}_{+1,+1}^{\rm{SO}}$ into the far-field mode ${\rm{LG}'}_{+1,+1}$.
That is, the scattering efficiency via the former path is written as 
\begin{widetext}
    \begin{eqnarray}
        (\sqrt{2ab\times\cos\theta})^2 \times \eta_{L0}&\times&(a\sqrt{\cos \theta})^2 \notag\\
        = (\sqrt{2ab\times\cos\theta})^2 &\times& \left(\frac{1}{2\pi}\frac{\Omega_0 \Omega'l^2\mu_B^2n}{32V_sc^2}\times\left(-f+2G_{44}\left(\frac{\mu_B}{2V_s}N\right)\right)^2\right)\times(a\sqrt{\cos \theta})^2 \label{eq:eff_0p1_p1p1_former}
    \end{eqnarray}
and calculated as $1.4\times10^{-22}$ by substituting the literature values and known values summarized in Table~\ref{table:known_values} into Eq.~(\ref{eq:eff_0p1_p1p1_former}). 
Note that while exit refraction generally projects an internal spin-orbit-coupled-light mode onto multiple far-field paraxial modes according to Eq.~(\ref{eq:LG_OptSOI}), throughout this paper we evaluate the conversion efficiency into the experimentally selected far-field paraxial mode.
Similarly, the scattering efficiency via the latter path is written as
    \begin{eqnarray}
        (\sqrt{b\times\cos\theta})^2 \times \eta_{0R}&\times&(\sqrt{2ab\times\cos \theta})^2 \notag\\
        = (b\sqrt{\cos\theta})^2 &\times& \left(\frac{1}{2\pi}\frac{\Omega_0 \Omega'l^2\mu_B^2n}{32V_sc^2}\times\left(-f-2G_{44}\left(\frac{\mu_B}{2V_s}N\right)\right)^2\right)\times(\sqrt{2ab\times\cos \theta})^2\label{eq:eff_0p1_p1p1_latter}
    \end{eqnarray}
\end{widetext}
and is calculated as $1.1\times10^{-28}$. The efficiency via former path is 5 orders of magnitude larger than that via latter path. The difference in efficiency between these two pathways comes from mainly the difference in the magnitude of $a$ and $b$. When the efficiency difference between these two pathways is very large, the larger one describes the scattering efficiency.

Similarly, the optical-vortex-anti-Stokes scattering from ${\rm{LG}}_{0,+1}$ to ${\rm{LG}'}_{-1,+1}$ shown as the significant blue bar with the gray arrow in Fig.~\ref{fig:eff_H}(a) arises dominantly from the transition from ${\rm{LG}}_{0,+1}^{\rm{SO}}$ to ${\rm{LG}'}_{0,0}^{\rm{SO}}$. 
The transition process from the input-far-field mode ${\rm{LG}}_{0,+1}$ to the scattered-optical-vortex-far-field mode ${\rm{LG}}'_{-1,+1}$ involves two pathways via BLS within the YIG sphere: one from ${\rm{LG}}_{0,+1}^{\rm{SO}}$ to ${\rm{LG}'}^{\rm{SO}}_{0,0}$ and another one from ${\rm{LG}}_{+1,0}^{\rm{SO}}$ to ${\rm{LG}'}_{+1,-1}^{\rm{SO}}$. For the former path, the initial spherical refraction first converts a $(a\sqrt{\cos \theta})^2$ of the far-field mode ${\rm{LG}}_{0,+1}$ into spin-orbit-coupled-light mode ${\rm{LG}}_{0,+1}^{\rm{SO}}$ in the dimension of power. This then undergoes BLS with an efficiency $\eta_{0L}$ into spin-orbit-coupled-light mode ${\rm{LG}'}_{0,0}^{\rm{SO}}$, and finally, spherical refraction converts a $(\sqrt{2ab\times\cos \theta})^2$ of the ${\rm{LG}'}_{0,0}^{\rm{SO}}$ into the far-field mode ${\rm{LG}'}_{-1,+1}$. That is, the scattering efficiency via the former path is written as 
\begin{widetext}
    \begin{eqnarray}
        (a\sqrt{\cos\theta})^2 \times \eta_{0L}&\times&(\sqrt{2ab\times\cos \theta})^2 \notag\\
        = (a\sqrt{\cos\theta})^2 &\times& \left(\frac{1}{2\pi}\frac{\Omega_0 \Omega'l^2\mu_B^2n}{32V_sc^2}\times\left(f-2G_{44}\left(\frac{\mu_B}{2V_s}N\right)\right)^2\right)\times(\sqrt{2ab\times\cos \theta})^2 \label{eq:eff_0p1_m1p1_former}
    \end{eqnarray}
\end{widetext}
and calculated as $1.4\times10^{-22}$. 
Equation~(\ref{eq:eff_0p1_m1p1_former}) is mathematically identical to Eq.~(\ref{eq:eff_0p1_p1p1_former}) for the Stokes process; consequently, it yields the same efficiency value.
Since the scattering efficiency of the latter path for the anti-Stokes process is mathematically identical to that in Eq.~(\ref{eq:eff_0p1_p1p1_latter}) for the Stokes process, and since that scattering efficiency is 5 orders of magnitude smaller than that of the former path described by Eq.~(\ref{eq:eff_0p1_m1p1_former}), Eq.~(\ref{eq:eff_0p1_m1p1_former}) describes the scattering efficiency of the anti-Stokes process.

\subsection{$R_i\rightarrow R_o$ configuration} \label{sec:details_theory_Ri2Ro}
We show the overall scattering efficiency of the optical-vortex-Stokes scattering shown as the red bar with the gray arrow in Fig.~\ref{fig:eff_H}(d). The transition process from the input-far-field mode ${\rm{LG}}_{0,-1}$ to the scattered-optical-vortex-far-field mode ${\rm{LG}}'_{+1,-1}$ involves two pathways via BLS within the YIG sphere: one from ${\rm{LG}}_{0,-1}^{\rm{SO}}$ to ${\rm{LG}'}^{\rm{SO}}_{0,0}$ and another one from ${\rm{LG}}_{-1,0}^{\rm{SO}}$ to ${\rm{LG}'}_{-1,+1}^{\rm{SO}}$. 
For the former path, the initial spherical refraction first converts a $(a\sqrt{\cos \theta})^2$ of the far-field mode ${\rm{LG}}_{0,-1}$ into spin-orbit-coupled-light mode ${\rm{LG}}_{0,-1}^{\rm{SO}}$ in the dimension of power. This then undergoes BLS with an efficiency $\eta_{0R}$ into spin-orbit-coupled-light mode ${\rm{LG}'}_{0,0}^{\rm{SO}}$, and finally, spherical refraction converts a $(\sqrt{2ab\times\cos \theta})^2$ of the ${\rm{LG}'}_{0,0}^{\rm{SO}}$ into the far-field mode ${\rm{LG}'}_{+1,-1}$. That is, the scattering efficiency via the former path is written as 
\begin{widetext}
    \begin{eqnarray}
        (a\sqrt{\cos\theta})^2 \times \eta_{0R}&\times&(\sqrt{2ab\times\cos \theta})^2 \notag\\
        = (a\sqrt{\cos\theta})^2 &\times& \left(\frac{1}{2\pi}\frac{\Omega_0 \Omega'l^2\mu_B^2n}{32V_sc^2}\times\left(-f-2G_{44}\left(\frac{\mu_B}{2V_s}N\right)\right)^2\right)\times(\sqrt{2ab\times\cos \theta})^2 \label{eq:eff_0m1_p1m1_former}
    \end{eqnarray}
and calculated as $0.12\times10^{-22}$ by substituting the literature values and known values summarized in Table~\ref{table:known_values} into Eq.~(\ref{eq:eff_0m1_p1m1_former}). Similarly, the scattering efficiency via the latter path is written as
    \begin{eqnarray}
        (\sqrt{2ab\times\cos\theta})^2 \times \eta_{L0}&\times&(b\sqrt{\cos \theta})^2 \notag\\
        = (\sqrt{2ab\times\cos\theta})^2 &\times& \left(\frac{1}{2\pi}\frac{\Omega_0 \Omega'l^2\mu_B^2n}{32V_sc^2}\times\left(-f+2G_{44}\left(\frac{\mu_B}{2V_s}N\right)\right)^2\right)\times(b\sqrt{\cos \theta})^2\label{eq:eff_0m1_p1m1_latter}
    \end{eqnarray}
\end{widetext}
and is calculated as $1.3\times10^{-27}$. 
The efficiency via former path is 4 orders of magnitude larger than that via latter path. 
Therefore, the efficiency of optical-vortex-Stokes scattering shown in Fig.~\ref{fig:eff_H}(d) is described by Eq.~(\ref{eq:eff_0m1_p1m1_former}).
Comparing Eq.~(\ref{eq:eff_0m1_p1m1_former}) with Eq.~(\ref{eq:eff_0p1_p1p1_former}) reveals that the relative sign between the Faraday ($f$) and Cotton-Mouton ($G_{44}$) magneto-optical coefficients is reversed, leading to destructive interference that suppresses the efficiency by $11\text{ dB}$.

Similarly, the optical-vortex-anti-Stokes scattering from ${\rm{LG}}_{0,-1}$ to ${\rm{LG}'}_{-1,-1}$ shown as the significant blue bar with the gray arrow in Fig.~\ref{fig:eff_H}(d) arises dominantly from the transition from ${\rm{LG}}_{-1,0}^{\rm{SO}}$ to ${\rm{LG}'}_{-1,-1}^{\rm{SO}}$.
Its scattering efficiency is mathematically identical to Eq.~(\ref{eq:eff_0m1_p1m1_former}) and calculated as $0.12\times10^{-22}$.

\subsection{$L_i\rightarrow R_o$ or $R_i\rightarrow L_o$ configuration} \label{sec:details_theory_Li2Ro_RiLo}
On the other hand, there are cases where the efficiencies of the two transition pathways are of the same order. These are the scatterings shown with gray arrows in Figs.~\ref{fig:eff_H}(b) and \ref{fig:eff_H}(c). In such cases, it is necessary to appropriately combine the probabilities of the two paths. Specifically, in the anti-Stokes scattering with the gray arrow in Fig.~\ref{fig:eff_H}(b), the transition process from the input far-field mode ${\rm{LG}}_{0,+1}$ to the scattered-optical-vortex-far-field mode ${\rm{LG}'}_{+1,-1}$ involves two pathways via BLS within the YIG sphere: one from ${\rm{LG}}^{\rm{SO}}_{+1,0}$ to ${\rm{LG}'}^{\rm{SO}}_{+1,-1}$ and another one from ${\rm{LG}}^{\rm{SO}}_{0,+1}$ to ${\rm{LG}'}^{\rm{SO}}_{0,0}$. In the expressions of efficiencies for both paths, the parts unrelated to magneto-optics are identical. Therefore, these efficiencies superimpose, and the overall scattering efficiency is given by
\begin{widetext}
    \begin{eqnarray}
        \frac{1}{2}&&\Big\{(\sqrt{2ab\times\cos\theta})^2 \times \eta_{R0}\times(a\sqrt{\cos \theta})^2\Big\} + \frac{1}{2}\Big\{(\sqrt{a\times\cos\theta})^2 \times \eta_{0L}\times(\sqrt{2ab\times\cos \theta})^2\Big\} \notag\\
        =&& \frac{1}{2}\Bigg\{(\sqrt{2ab\times\cos\theta})^2 \times \left(\frac{1}{2\pi}\frac{\Omega_0 \Omega'l^2\mu_B^2n}{32V_sc^2}\times\left(f+2G_{44}\left(\frac{\mu_B}{2V_s}N\right)\right)^2\right)\times(a\sqrt{\cos \theta})^2\Bigg\} \notag\\
         &&+\frac{1}{2}\Bigg\{(a\sqrt{\cos\theta})^2 \times \left(\frac{1}{2\pi}\frac{\Omega_0 \Omega'l^2\mu_B^2n}{32V_sc^2}\times\left(f-2G_{44}\left(\frac{\mu_B}{2V_s}N\right)\right)^2\right)\times(\sqrt{2ab\times\cos \theta})^2\Bigg\}
        \label{eq:eff_0p1_p1m1_all}        
    \end{eqnarray}
\end{widetext}
and calculated as $0.78\times10^{-22}$. The leading $1/2$ in both terms reflects the fact that the excited magnons contribute nearly equally to each path. 
Here, we state the reason why the summation of two scattering efficiencies, as shown in Eq.~(\ref{eq:eff_0p1_p1m1_all}), is appropriate. While each individual scattering path is a coherent process and we have confirmed that the observed scattered modes in the far field are also coherent, the overall scattering efficiency does not exhibit quantum interference and is treated as a simple summation of the efficiencies from two different pathways. This is because the initial and final states of the spin-orbit-coupled modes inside the sphere involved in these two paths are distinct and orthogonal to each other. According to the fundamental principles of quantum mechanics, when pathways are in principle distinguishable, quantum interference vanishes \cite{RR1965}. The validity of this physical interpretation is further supported by the excellent quantitative agreement between the experimental results and the theoretical calculations shown in Table~\ref{table:comparison_exp_theo}.

Similarly, in the Stokes scattering with the gray arrow in Fig.~\ref{fig:eff_H}(c), the transition process from the input far-field mode ${\rm{LG}}_{0,+1}$ to the scattered-optical-vortex-far-field mode ${\rm{LG}'}_{-1,+1}$ involves two pathways via BLS within the YIG sphere: one from ${\rm{LG}}^{\rm{SO}}_{-1,0}$ to ${\rm{LG}'}^{\rm{SO}}_{-1,+1}$ and another one from ${\rm{LG}}^{\rm{SO}}_{0,-1}$ to ${\rm{LG}'}^{\rm{SO}}_{0,0}$. 
These scattering efficiencies superimpose. The overall scattering efficiency is mathematically identical to Eq.~(\ref{eq:eff_0p1_p1m1_all}) and calculated as $0.78\times10^{-22}$.

Here, Table~\ref{table:comparison_exp_theo} summarizes the calculated and experimental efficiencies for all scatterings observed in Figs.~\ref{fig:eff_H}(a)-\ref{fig:eff_H}(d) that are labeled with gray arrows.

\section{Calculated scattering efficiency for optical vortex generation with $l_p=-3$} \label{sec:lp=m3}
We consider why scattering with the combination $(\Delta s_m,\Delta s_p,\Delta l_p)=(+1,+2,-3)$ satisfying the conservation of total angular momentum between light and magnons, as described in Appendix~\ref{sec:lp=3}, is not observed. From Eq.~(\ref{eq:LG_OptSOI}), writing down the correspondence between the far-field mode and the spin-orbit-coupled-light modes inside the sphere for the input-Gaussian-paraxial-light mode and the scattered-paraxial-vortex-light mode yields
\begin{widetext}
\begin{equation}
    {\rm{LG}}_{0,-1}=\sqrt{\cos\theta}\left(a\times{\rm{LG}}_{0,-1}^{\rm{SO}}-b\times{\rm{LG}}_{-2,+1}^{\rm{SO}}-\sqrt{2ab}\times{\rm{LG}}_{-1,0}^{\rm{SO}}\right) \label{eq:LG0m1_OptSOI}
\end{equation}
and
\begin{equation}
 \sqrt{\cos\theta}\left(a\times{\rm{LG}'}_{-3,+1}^{\rm{SO}}-b\times{\rm{LG}'}_{-1,-1}^{\rm{SO}}-\sqrt{2ab}\times{\rm{LG}'}_{-2,0}^{\rm{SO}}\right)={\rm{LG}'}_{-3,+1}. \label{eq:LGm3p1_OptSOI}
\end{equation}
Following the procedure outlined in Appendix~\ref{sec:details_theory}, we can expect anti-Stokes scattering to occur from ${\rm{LG}}_{-2,+1}^{\rm{SO}}$ to ${\rm{LG}'}^{\rm{SO}}_{-2,0}$ and from ${\rm{LG}}^{\rm{SO}}_{-1,0}$ to ${\rm{LG}'}^{\rm{SO}}_{-1,-1}$, which share the same OAM. The scattering efficiency is denoted as
    \begin{eqnarray}
        \frac{1}{2}&&\Big\{(b\sqrt{\cos\theta})^2 \times \eta_{0L}\times(\sqrt{2ab\times\cos \theta})^2\Big\} + \frac{1}{2}\Big\{(\sqrt{2ab\times\cos\theta})^2 \times \eta_{R0}\times(b\sqrt{\cos \theta})^2\Big\} \notag\\
        =&& \frac{1}{2}\Bigg\{(b\sqrt{\cos\theta})^2 \times \left(\frac{1}{2\pi}\frac{\Omega_0 \Omega'l^2\mu_B^2n}{32V_sc^2}\times\left(f+2G_{44}\left(\frac{\mu_B}{2V_s}N\right)\right)^2\right)\times(\sqrt{2ab\times\cos \theta})^2\Bigg\} \notag\\
         &&+\frac{1}{2}\Bigg\{(\sqrt{2ab\times\cos\theta})^2 \times \left(\frac{1}{2\pi}\frac{\Omega_0 \Omega'l^2\mu_B^2n}{32V_sc^2}\times\left(f-2G_{44}\left(\frac{\mu_B}{2V_s}N\right)\right)^2\right)\times(b\sqrt{\cos \theta})^2\Bigg\}
        \label{eq:eff_0m1_m3p1_all}        
    \end{eqnarray}
and is calculated $7.1\times10^{-28}$ using the literature and known values in Table~\ref{table:known_values}. We can conclude that the reason it is not measured this time is simply its low efficiency.

\end{widetext}

\begin{table}[t]
  \begin{center}
    \caption{Literature and known values.} 
    \begin{tabular}{|c|c|} \hline
      Physical quantity & Value   \\ \hline \hline 
       Aperture angle: $\theta$ &$0.11\,\rm{rad}$ \\ \hline 
       \begin{tabular}{c} Angular frequency of\\ incident light: $\Omega_0$ \end{tabular} & $2\pi\times193\,\rm{THz}$\\ \hline
       \begin{tabular}{c} Angular frequency of\\ scattered light: $\Omega'$ \end{tabular} & $2\pi\times(193\pm0.00373)\,\rm{THz}$\\ \hline
       Interaction length: $l$ &$0.5\,\rm{mm}$ \\ \hline 
       Spin density of YIG~\cite{DA2009}: $n$ &$2.1\times10^{28}\,\rm{m^{-3}}$ \\ \hline 
       Sample volume: $V_s$ &$\displaystyle \frac{4}{3}\pi\left(\frac{l}{2}\right)^3$ \\ \hline 
        \begin{tabular}{c} Faraday coefficient of YIG\\ for $1550\,\rm{nm}$ light~\cite{D1991} : $f$ \end{tabular} & $\displaystyle \frac{2\sqrt{\epsilon_r}\nu}{k_0(-\frac{1}{2}\mu_B n)}$\\ \hline
        \begin{tabular}{c} Relative permittivity of YIG\\ for $1550\,\rm{nm}$ light~\cite{DA2009}: $\epsilon_r$ \end{tabular} & $2.2$\\ \hline
        Verdet constant of YIG~\cite{MS2000}: $\nu$ &$380\,\rm{rad/m}$ \\ \hline 
        \begin{tabular}{c} Cotton-Mouton coefficient of YIG\\ for $1550\,\rm{nm}$ light~\cite{DA2009}: \\ $G_{44}\left(\frac{\mu_B}{2}n\right)^2$ \end{tabular} & $-1.14\times10^{-4}$\\ \hline
    \end{tabular}
  \label{table:known_values}
  \end{center}
\end{table}

\begin{table}[H]
  \begin{center}
    \caption{Experimentally observed and calculated scattering efficiencies.} 
    \begin{tabular}{|c|c|c|} \hline
      Scattering & Experimental & Calculated   \\ \hline \hline 
       Stokes in Fig.~\ref{fig:eff_H}(a) & $0.85\times10^{-22}$ & $1.4\times 10^{-22}$ \\ \hline 
       Anti-Stokes in Fig.~\ref{fig:eff_H}(a) & $0.89\times10^{-22}$ & $1.4\times 10^{-22}$ \\ \hline 
        Anti-Stokes in Fig.~\ref{fig:eff_H}(b) & $0.53\times10^{-22}$ & $0.78\times 10^{-22}$ \\ \hline 
        Stokes in Fig.~\ref{fig:eff_H}(c) & $0.57\times10^{-22}$ & $0.78\times 10^{-22}$ \\ \hline
        Stokes in Fig.~\ref{fig:eff_H}(d) & $0.07\times10^{-22}$ & $0.12\times 10^{-22}$ \\ \hline
        Anti-Stokes in Fig.~\ref{fig:eff_H}(d) & $0.07\times10^{-22}$ & $0.12\times 10^{-22}$ \\ \hline
    \end{tabular}
  \label{table:comparison_exp_theo}
  \end{center}
\end{table}

\newpage









\end{document}